\documentclass[aip,reprint,amsmath,amssymb]{revtex4-1}
\usepackage[utf8]{inputenc}
\usepackage{graphicx}
\usepackage{bm}
\usepackage{amsfonts,amssymb,amsmath}
\usepackage{color}
\usepackage{comment} 
\usepackage{wasysym} 
\usepackage{chemformula} 
\usepackage{textcomp}
\usepackage[french,english]{babel}
\usepackage[hang,nooneline]{subfigure}
\usepackage{pdfpages}
\usepackage{pgffor}
\usepackage{hyperref}
\makeatletter
\AtBeginDocument{\let\LS@rot\@undefined}
\makeatother

\begin{document}
\title{Probing the concept of line tension down to the nanoscale}
\author{Romain Bey}
\affiliation{Univ. Grenoble Alpes, CNRS, LIPhy, 38000 Grenoble, France}
\author{Benoit Coasne}\email{benoit.coasne@univ-grenoble-alpes.fr}
\affiliation{Univ. Grenoble Alpes, CNRS, LIPhy, 38000 Grenoble, France}
\author{Cyril Picard}\email{cyril.picard@univ-grenoble-alpes.fr}
\affiliation{Univ. Grenoble Alpes, CNRS, LIPhy, 38000 Grenoble, France}
\date{\today}

\begin{abstract}
A novel mechanical approach is developed to explore by means of atom-scale simulation the concept of line tension at a solid-liquid-vapor contact line as well as its dependence on temperature, confinement, and solid/fluid interactions. More precisely, by estimating the stresses exerted along and normal to a straight contact line formed within a partially wet pore, the line tension can be estimated while avoiding the pitfalls inherent to the geometrical scaling methodology based on hemispherical drops. The line tension for Lennard-Jones fluids is found to follow a generic behavior with temperature and chemical potential effects that are all 
included in a simple contact angle parameterization.  
Former discrepancies between theoretical modeling and molecular simulation are resolved, and the line tension concept is shown to be robust down to molecular confinements. The same qualitative behavior is observed for water but the line tension at the wetting transition diverges or converges towards a finite value depending on the range of the solid/fluid interactions at play.

\end{abstract}

\maketitle

\section{Introduction}

The contact line between three phases is a particular \emph{locus} conducive to physical couplings between the macroscopic and the molecular scales. Diverse phenomena are  intrinsically bound to the presence of a contact line, such as heterogeneous nucleation \cite{guillemot_activated_2012,remsing_pathways_2015, Tinti2017}, formation of nanovesicles from a membrane \cite{Satarifard2018}, dynamical wetting \cite{dG1985,Bonn2009,Lhermerout2019}, and as a last example among others stabilisation and pinning of nanoinclusions at an interface (bubbles, droplets, colloids, etc.) \cite{lohse_surface_2015, bresme_computer_1998, Tan2017}. To better understand these phenomena, which play a key role in several applicative fields such as biotechnology (nanoemulsion, encapsulation) \cite{Singh2017}, chemical engineering (catalysis, electrochemistry) \cite{Debe2012} or process engineering (boiling, condensation) \cite{Karayiannis2017}, a challenge remains in deciphering the effects due to the presence of a contact line. At the micron scale,  the wedge that is formed in the vicinity of a solid/liquid/vapor contact line may be responsible for enhanced heat and mass transfer at the origin for instance of the well known coffee ring effect \cite{Deegan1997}. At the nanometer scale, such a wedge has been identified to give rise to specific molecular interactions close to the three-phase contact line, at the origin of a specific free energy contribution: the line tension \cite{joanny_role_1986, Drelich1996, weijs_origin_2011}.  This specific thermodynamical quantity was first mentioned by Gibbs \cite{gibbs_1876}, who introduced the line tension $\tau$ as an excess free energy per unit length of contact line or a tangential force along this line. Unlike surface tension of planar fluid/fluid interfaces, line tension can be either positive or negative as there is no thermodynamical argument to predict its sign \cite{rowlinson}. From simple scaling arguments, the order of magnitude of $\tau$ for water is expected to be $\left|\tau\right  | \sim \gamma_{lv} \sigma\sim 20$~pN with $\gamma_{lv}$ the liquid/vapor surface tension and $\sigma$ the molecular size \cite{guillemot_activated_2012}.  Such a scaling implicitly suggests that the impact of line tension is limited to the molecular scale \cite{Drelich1996}. Nevertheless, in the case of nanostructured materials, the effect of line tension can be scaled up and impact macroscopic phenomena. For instance, it has been recently shown that line tension could control the macroscopic bulk pressure required to induce capillary drying within hydrophobic nanoporous material \cite{guillemot_activated_2012,Tinti2017}.

Despite its wide range of possible contributions, the understanding of line tension is still limited.  Available measurements, either from experiments \cite{Wang1999,Pompe2000,checco_nonlinear_2003,kameda2008,berg_impact_2010, mcbride2012,Heim2013} or based on numerical simulations \cite{werder2003,Hirvi2006,winter2009,weijs_origin_2011,zhang_influence_2014}, lead to mostly negative but also positive values for the line tension with magnitude spanning from $10^{-6}$ to $10^{-12}$~N, this dispersion being probably due to the diversity of methods and systems under investigation \cite{law2017}. Moreover, from analytical predictions of Joanny and De Gennes, a divergence of line tension towards infinite positive values is expected for Lennard-Jones fluids at the wetting transition \cite{joanny_role_1986}. From this variability of $\tau$, one must admit that estimating accurately line tension remains a delicate task with long-standing debates about the role of competing effects, such as line pinning \cite{checco_nonlinear_2003} or surface curvature corrections \cite{Joswiak2013,kanduc_going_2017,das_contact_nodate}. Up to date, no consensus has been met on the dependence of line tension on physical parameters such as temperature, substrate hydrophilicity or fluid molecular structure.

Most experimental and molecular simulation  measurements of line tension consist of evaluating the dependence of the contact angle $\theta$ of a sessile drop on the radius $r$ of its circular contact line. In this approach, a simple decomposition into surface and line free energies reveals that a geometrical scaling, known as the modified Young equation, is expected \cite{boruvka_generalization_1977}:
\begin{equation}
\cos\theta = \cos\theta_{Y}-\frac{\tau}{\gamma_{lv}r}
\label{Eq1}
\end{equation}
where $\cos \theta_{Y} = (\gamma_{sv}-\gamma_{sl})/\gamma_{lv}$ is the Young contact angle defined from the solid-vapor $\gamma_{sv}$, solid-liquid $\gamma_{sl}$ and liquid-vapor $\gamma_{lv}$ surface tensions. Eq. [\ref{Eq1}] implicitly assumes that surface and line tensions only depend on the nature of the materials so that changes in $\cos \theta$ are proportional to $1/r$. However, surface and line tensions can vary with the fluid chemical potential as well as with surface and line curvatures \cite{boruvka_generalization_1977, ward_effect_2008}. In this respect, while curvature seems to have a negligible impact on $\gamma_{lv}$ (expressed with the so-called Tolman length) down to submolecular sizes, its effect on line tension remains to be established \cite{ das_contact_nodate, kanduc_going_2017}. Moreover, the geometrical approach in Eq. [\ref{Eq1}] suffers from difficulties in evaluating 
the shape of the sessile drop. In particular, experimental departures from Eq. [\ref{Eq1}] \cite{checco_nonlinear_2003,berg_impact_2010} cast doubt on the relevance of the geometrical scaling methodology to measure line tensions $\tau$ \cite{marmur_line_1997,ward_effect_2008,das_contact_nodate}. Even in molecular simulation approaches, the position of interfaces and lines is ambiguous at the molecular scale 
so that line tensions estimated numerically are subjected to large error bars.

\begin{figure}[ht]
\centering
\includegraphics[width=8.6cm, keepaspectratio=true]{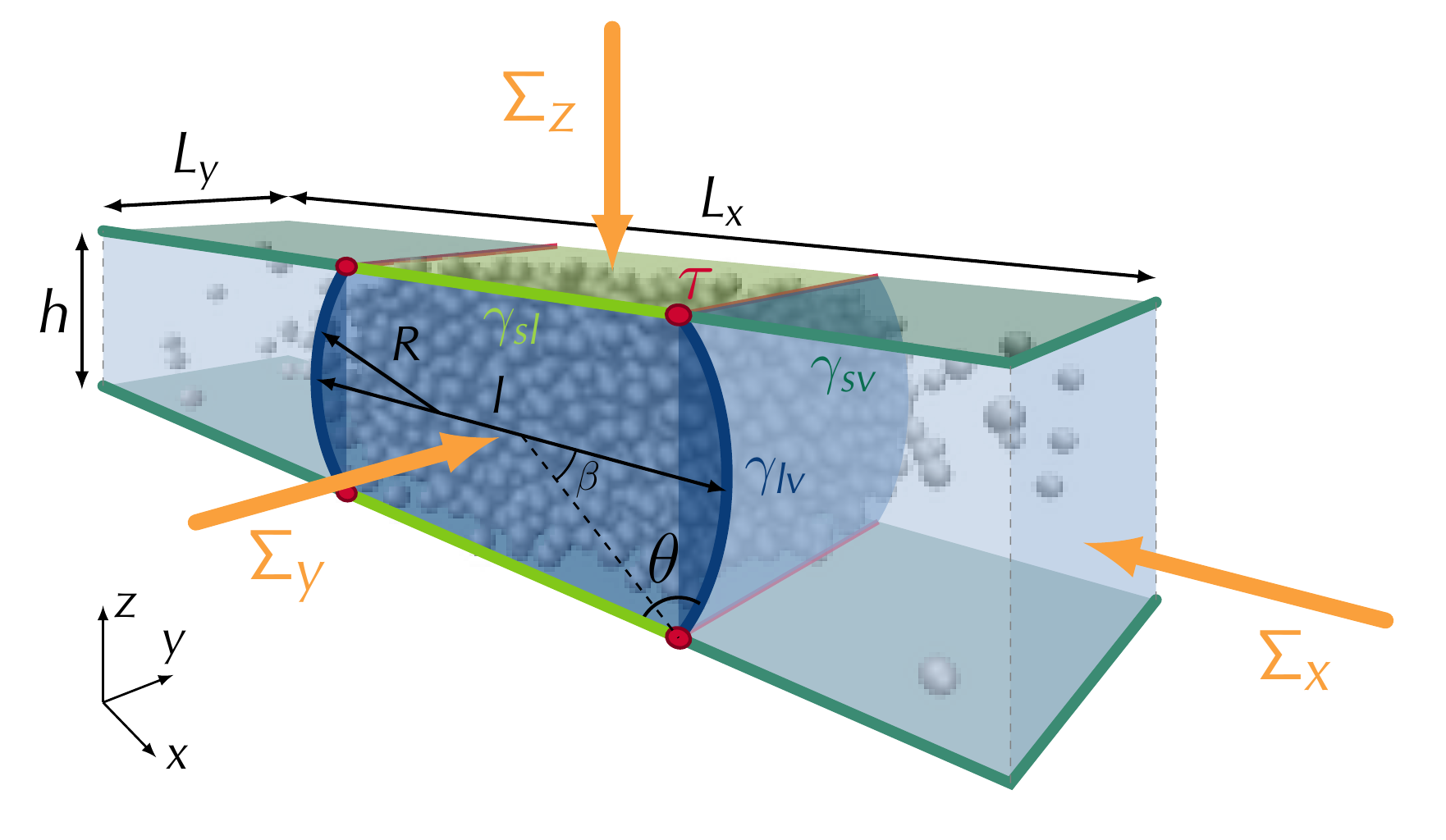}
\caption{Set-up consisting of a liquid in contact with its vapor confined between two solid walls (perpendicular to the $z$ direction). {Solid/fluid interfaces are} separated by a distance $h$. Periodic boundary conditions are applied along the $x$ and $y$ directions which are parallel and perpendicular to the red contact lines. The line tension $\tau$ {of these straight contact lines} is estimated from the forces $\Sigma_{x}$, $\Sigma_{y}$ and $\Sigma_{z}$ exerted along $x$, $y$ and $z$.}
\label{Fig1}
\end{figure}

In this paper, a molecular simulation methodology is developed to estimate line tensions $\tau$ without geometrical scaling. The effects of temperature, confinement, solid/fluid interactions, and chemical potential are investigated for any dispersive fluid modeled by a Lennard-Jones potential (LJ) and for water. In the spirit of the seminal work of Tarazona \cite{Tarazona1981} and recent approach of Shao et al \cite{shao_anisotropy_2015}, statistical mechanics expressions are used to determine $\tau$ from the stress anisotropy in the vicinity of a triple line. {While the approach of Shao et al \cite{shao_anisotropy_2015} is mostly applicable to three fluid systems with limited interface curvature, our approach addresses the case of two fluids confined between solid walls (Fig. \ref{Fig1}) without any restriction on the curvature of the fluid/fluid interface, that is for any contact angle and any confinement}. In this approach, a liquid in contact with its vapor is confined, at a temperature $T$, in a slit pore of a width $h$ formed by two solid planar walls perpendicular to the $z$ direction (Fig. \ref{Fig1}). The system is infinite in the $x$ and $y$ directions thanks to the use of periodic boundary conditions (see methodological details in \emph{Materials and Methods} and \emph{Supplementary Information}). 

Our approach  avoids the delicate computation of local stresses or pressures at the vicinity of the contact line. The value of $\tau$ is directly extracted from the measurement of total forces $\Sigma_{x}$, $\Sigma_{y}$ and $\Sigma_{z}$ in the three space directions. This approach is used to probe the line tension of the contact line of a Lennard-Jones fluid or water within various confinements.

\section{Material and methods}

\subsection{A mechanical route to measure line tension}

\begin{figure}[ht]
\centering
\subfigure[]{\label{Kp}\includegraphics[width=0.49\linewidth, keepaspectratio=true]{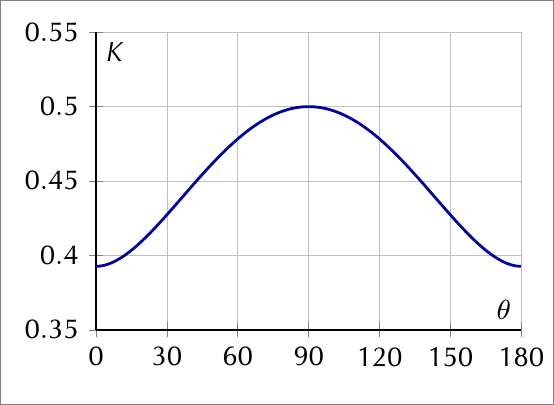}}\hfill\subfigure[]{\label{HBp}\includegraphics[width=0.49\linewidth, keepaspectratio=true]{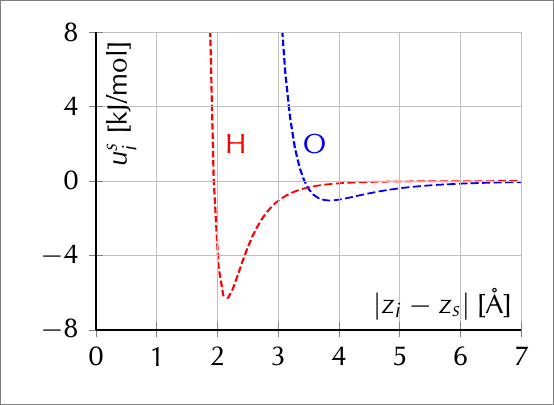}}
\caption{\subref{Kp} Trigonometric function $K(\theta)$ accounting for bulk and surface contributions to the meniscus free energy $\Sigma_{m}$. \subref{HBp} Interaction potentials of the hydrogen-bonding solid used in the water simulations. The hydrogen of water is strongly attracted towards the solid, therefore mimicking a hydrogen bond interaction.}
\label{SI.fig}
\end{figure}

At mechanical equilibrium the total force $\Sigma_x$ exerted on the fluid system through a $yz$ plane does not depend on the $x$ location of the plane. For a plane within the vapor phase far from the liquid phase, $\Sigma_x$ relies on the vapor pressure $P_v$  and the solid-vapor surface tension $\gamma_{sv}$ while for a plane within the liquid phase far from the vapor phase it relies on the liquid pressure $P_l$  and the solid-liquid surface tension $\gamma_{sl}$:
\begin{equation}
\Sigma_{x}=L_{y}(-P_{v}h+2\gamma_{sv})=L_{y}(-P_{l}h+2\gamma_{sl}) 
\label{Eq2}
\end{equation}
Due to translational invariance the force $\Sigma_{y}$ exerted on a slice of fluid (vapor + liquid) in the $y$ direction does not depend on the location of the cutting plane. It arises from the three surface tensions $\gamma_{lv}$, $\gamma_{sl}$ and $\gamma_{sv}$, the liquid and vapor pressures $P_{l}$ and $P_{v}$, and the line {tension $\tau$:
\begin{equation}
\Sigma_{y}=-P_{v}A_{v}-P_{l}A_{l} +l_{sv}\gamma_{sv}+l_{sl}\gamma_{sl}+l_{lv}\gamma_{lv}+4\tau
\label{Eq3}
\end{equation}
where $A_{v}$ and $A_{l}$ are the surface areas of the vapor and liquid phases in $xz$ plane, $l_{sv}$, $l_{sl}$ and $l_{lv}$ the lengths corresponding to the intersections of respectively solid/vapor, solid/liquid and liquid/vapor interfaces with $xz$ plane. To facilitate the determination of $\tau$ from Eq. [\ref{Eq3}] we introduce} the algebraic area $A_{m}$ of each meniscus region (dark blue regions in Fig. \ref{Fig1}), from which the liquid and vapor areas write respectively $A_{l}=hl_{sl}/2+2A_{m}$ and $A_{v}=h(L_{x}-l_{sl}/2)-2A_{m}$. The algebraic area $A_{m}$, of the same sign as $P_{l}-P_{v}$, is positive for contact angle $\theta >\pi/2$ and negative for $\theta <\pi/2$. Replacing $A_{l}$ and $A_{v}$ by their respective expressions and using the mechanical balance in the $x$ direction Eq. [\ref{Eq2}],  the force in the $y$ direction simplifies without the solid/fluid surface tensions :

\begin{equation}
\Sigma_{y}=\frac{L_{x}}{L_{y}}\Sigma_{x}
- 2  A_{m}(P_{l}-P_{v})  +l_{lv} \gamma_{lv}+4 \tau
\label{SI.equation8}
\end{equation}
As $P_{l}$, $P_{v}$ and $\gamma_{lv}$ are considered as uniform, the radius of curvature $R$, of the liquid/vapor interface, is constant. The role of the line tension $\tau$ is precisely to take into account the molecular interactions between the three phases at the vicinity of the contact line that may alter locally this uniformity \cite{joanny_role_1986}. 
From the circular geometry of the liquid vapor interface, $A_{m}={R^{2}}( \beta -\sin(2 \beta)/2)$ and $l_{lv}/2=2R\beta$ with $\beta=\theta-\pi/2$ standing for half the angle of the arc formed by a cut in a zx plan of the confined liquid/vapor interface (blue lines in Fig. \ref{Fig1}). In these expressions the radius of curvature $R={h}/({2 \sin \beta})$ is an algebraic quantity the same sign as $\beta$. Replacing $A_{m}$, $l_{lv}$ and $R$ by their expressions and using Laplace's law of capillarity, $P_{l}-P_{v}=2\gamma_{lv}/R$, the force $\Sigma_{y}$ further simplifies without pressures:
\begin{equation}
\Sigma_{y}=\frac{L_{x}}{L_{y}}\Sigma_{x}+ \gamma_{lv}h \frac{\sin(2 \beta)+2\beta}{ 2\sin(\beta)}+ 4 \tau
\label{SI.equation11}
\end{equation}
Isolating $\tau$ from Eq. [\ref{SI.equation11}] leads to the central relation of the paper:
\begin{equation}
\tau=\frac{1}{2}\Sigma_{m}-\gamma_{lv}hK(\theta)
\label{Eq4} 
\end{equation}
where the force $\Sigma_{m}=(\Sigma_{y}-\Sigma_{x}L_{x}/L_{y})/2$ corresponds to the total force in the $y$ direction acting on each meniscus region ({dark blue regions, blue lines and red dots} in Fig. \ref{Fig1}). The function $K(\theta)$ is a combination  of trigonometric functions, which weakly depends on the contact angle (see  Fig. \ref{Kp}):
\begin{equation}
K(\theta)=\frac{\sin(2\beta)+2 \beta}{8 \sin(\beta)}=\frac{1}{4}\Big(\sin(\theta)-\frac{\theta-\pi/2}{\cos(\theta)}\Big)
\label{SI.equation12}
\end{equation}
The function K is symmetric with respect to the angle $\theta=\pi/2$ (or $\beta=0$). The line tension {$\tau$ given by Eq. [\ref{Eq4}] is thus obtained from the difference of half the total force applied} on a meniscus region in the $y$ direction, that is the total force $\Sigma_{m}/2$ based on line, surface and bulk contributions,  minus the term $\gamma_{lv}hK(\theta)$ which stands for the force due to bulk and surfaces only.

The forces $\Sigma_{x}$ and $\Sigma_{y}$ are computed using the virial expression of anisotropic stresses at an unstructured solid surface from the positions of the $N$ fluid particles \cite{schofield1982,thompson_general_2009}:  

\begin{equation}
\Sigma_{\alpha}=\Big\langle - \frac{Nk_{B}T}{L_\alpha}+W_\alpha\Big\rangle
\label{Eq7}
\end{equation}
with $\alpha=x,y$ and $k_{B}$  Boltzmann's constant. $W_\alpha$ is the energy derivative relative to a homogeneous affine expansion of all the {fluid} atomic positions and the system boundaries in the direction $\alpha$.

Unlike the forces $\Sigma_{x}$, $\Sigma_{y}$ and $\Sigma_{m}$, line tension $\tau$ depends on the geometrical parameterization of the system. This dependence underlines that line tension is not an intrinsic parameter \cite{schimmele_conceptual_2007} except {in the special case} of a straight triple line formed by three fluids at the same pressure \cite{rowlinson}. This 1D {exception} is analogous to the 2D case of the intrinsic surface tension characterizing a planar interface that separates two phases at the same pressure. In the absence of pressure uniformity (as encountered here), line and surface tensions depend on the definition of interface position \cite{ward_effect_2008}. This is the case if interfaces are curved or in the presence of a solid phase under non-isotropic stress for which the concept of scalar pressure must be replaced by an elastic stress tensor. Surface and line tensions are  therefore  dependent on the choice made to define the position of the surface and contact angle {as it appears in Eq. [\ref{Eq4}] where the value of $\tau$ depends on the geometrical quantities $h$ and $\theta$ that may be defined according to various conventions at the molecular scale}.
{To adopt a definition of the distance $h$ that is based on physical parameters, we consider the surface excess $\Gamma_{l}$ of fluid at the solid-liquid interface. The conservation of fluid mass in a vertical slab located in the liquid phase writes $2\Gamma_{l} + h\rho_{l} = n_{l}$ with $\rho_{l}$ the bulk liquid density (see \emph{Supplementary Information} and Fig. \ref{Fig2}) and $n_{l}$ the average number of molecules per unit of solid surface area in a slab perpendicular to the $x$ direction taken within the liquid (inset of Fig. \ref{Fig2}).  Using the Gibbs convention, that is a zero fluid adsorption $\Gamma_{l} = 0$, leads to the definition of $h$ that verifies $h = n_{l}/\rho_{l}$. 
The second geometric parameter, the contact angle $\theta$, is then deduced from Laplace's law of capillarity:}
\begin{equation}
\cos \theta = -\frac{(P_{l}-P_{v})h}{2\gamma_{lv}}= \frac{h}{l}\Big[ \frac{1}{2\gamma_{lv}L_{y}}\Big(\Sigma_{z}+P_{v}L_{x}L_{y}\Big)-1\Big]
\label{Eq9}
\end{equation}
where $\Sigma_{z}$, the force exerted by the fluid on the bottom solid surface, subdivides into volume and surface contributions $\Sigma_{z}~=~L_{y}[2 \gamma_{lv} -P_{l}l -P_{v}(L_{x}-l)]$. 
This definition is self consistent with the establishment of Eqs. [\ref{Eq4}] and [\ref{SI.equation12}]. In the end, the numerical values of $\gamma_{sl}$ and $\gamma_{sv}$ are not needed to extract $\tau$ as these quantities simplify in the derivation. 
The distance $l$ between the two vapor/liquid menisci is defined through the zero adsorption criterion at the liquid-vapor interface: $l= [n_{m}-\rho_{v}L_{x}]/[\rho_{l}-\rho_{v}]$ where $n_{m}$ is the average number of fluid molecules per unit of surface area in a slab perpendicular to the $z$ direction at the center of the slit (inset Fig. \ref{Fig2}). The vapor pressure $P_v$ is determined from the mean vapor density $\rho_{v}$ in the middle of the dry pore region using the ideal gas law, $P_{v}=\rho_{v} k_{\textrm{B}}T$. 
{Liquid-vapor surface tension $\gamma_{lv}$ is measured in an independent simulation of an infinite planar liquid-vapor interface following the classical mechanical route (see Supplementary Information).
The definition Eq. [\ref{Eq9}]} of the contact angle was used as it is relevant for any confinement $h$ and solid-fluid interaction $\varepsilon_{s}$.

\subsection{Lennard-Jones fluid}

The Lennard-Jones fluid involves repulsive and dispersive interactions with a total energy given by
\begin{equation}
U=\sum_{i<j}u_{ij}^{f}+\sum_{i}\sum_s u_{i}^{s}
\end{equation}
where $u_{ij}^{f}$ is the pair additive fluid interaction
\begin{equation}
u_{ij}^{f}=4\varepsilon\Big[\Big(\frac{\sigma}{r_{ij}}\Big)^{12}-\Big(\frac{\sigma}{r_{ij}}\Big)^{6}\Big]
\label{Eq11}
\end{equation}
and $u_{i}^{s}$ is the fluid-wall interaction
\begin{equation}
u_{i}^{s}=\varepsilon_{s}\Big[\frac{2}{15}\Big(\frac{\sigma_{s}}{\vert z_{i}-z_{s}\vert}\Big)^{9}-\Big(\frac{\sigma_{s}}{\vert z_{i}-z_{s}\vert}\Big)^{3}\Big]
\label{Eq12}
\end{equation} 
{with $s$ referring to the lower or upper solid plates located at $z_{s}$ ($s = l,u$), $r_{ij}$ the distance between fluid particles $i$ and $j$, $z_{i}$ the z-position of fluid particle $i$ and  $(\varepsilon, \sigma)$ and $(\varepsilon_s, \sigma_s)$ respectively the LJ fluid-fluid and solid-fluid interaction parameters (unless stated otherwise, $\sigma_{s}=2\sigma$). The fluid-wall 9-3 interaction Eq. [\ref{Eq12}] corresponds to the regular LJ potential integrated over a half space infinite solid. A truncation cutoff radius of $r_{c}=4 \sigma$ is used for fluid-fluid interactions. The temperature $T$ is constrained using a Nos\' e-Hoover thermostat with a damping time $t_{\text{damp}}=100\Delta t$.}
In our simulation, the number of particles $N$ varies between 560 and 18900. Each molecular dynamics simulation runs at constant volume $V$ adjusted according to the number of particles, $42\sigma<L_{x}<90\sigma$, $L_{y}=21\sigma$ and $L_{z}=80\sigma$. The simulations consist of $10^{8}$ timesteps with $\Delta t=0.005\, \sigma\sqrt{m/\varepsilon}$  ($m$ is the mass of the particle). The forces $\Sigma_\alpha$ ($\alpha=x,y$) expressed in Eq. [\ref{Eq7}] of the main article are computed every $100\Delta t$. For post analysis, the configurations are stored every $10^4 \Delta t$. Error bars are computed using the block averaging method on {blocks of size} $9\times 10^{6}\Delta t$, and are of the order of the symbol size in Fig. \ref{Fig3}.

\subsection{Water}

Water molecular dynamics simulations were carried out with 480 to 3960 rigid SPC/E molecules in simulation boxes of dimensions
\begin{align*} 
L_{x}&=[12,24]~\textrm{nm}\\
L_{y}&=4~\textrm{nm}\\
L_{z}&=[3.35,3.45,3.55,3.65,3.75,3.85,4.85,6.15]~\textrm{nm}
\end{align*} 
The confinement parameter $h$ defined through the zero fluid  adsorption condition at the solid-liquid surface varies between $1$ and $5$~nm. The line tensions shown in Fig. [\ref{Fig5}] correspond to the case $L_z=3.75$~nm and $h=2.6$~nm. In the SPC/E model, fluid-fluid dispersive interactions are modeled using a Lennard-Jones potential with a cutoff radius of $9$ \AA. Electrostatic forces are computed using a cutoff radius of 9 $\textrm{\AA}$  coupled to a long-range correction computed through the PPPM algorithm. To avoid interactions between periodic images in the $z$ direction, the methodology exposed in \cite{yeh_ewald_1999} is applied, inserting 3 empty boxes between the periodic images in $z$ direction.
 The motion of the water molecules is integrated using a rigid-body integrator with a timestep of $\Delta t =2$ fs. The temperature is constrained to $T=300$~K using a Nos\'e-Hoover thermostat with a characteristic damping  time $t_{damp}=1$ ps. Configurations are stored every $2$ ps for post-analysis. Simulations are run for a total duration of $50$ ns. Error bars are computed using the block averaging method on blocks of $450$~ps.

The interaction between water and the solid is either chosen to be dispersive or to involve hydrogen-bonding. In the dispersive case, only oxygen atoms interact with the solid. To model this interaction, using Eq. [\ref{Eq12}], we choose  $\sigma_{s}=3$ $\textrm{\AA}$ and $\varepsilon_{s}$ is varied between $1.57$ and $7.22$ kJ/mol to scan a broad range of hydrophilicities. In the hydrogen-bonding case, we model the water-solid interactions through a potential:\begin{equation}
u^{s}_i=\frac{\eta \varepsilon_{s}}{n-m}\Big[m\Big(\frac{\sigma_{s}}{\vert z_i-z_s \vert}\Big)^{n}-n\Big(\frac{\sigma_{s}}{\vert z_i-z_s \vert}\Big)^{m}\Big]
\label{equation29}
\end{equation}
with $[n,m]=[12,6]$, $\sigma_{s}=3.85$ $\textrm{\AA}$ and $\varepsilon_{s}=1.05$ kJ/mol for oxygen atoms and $[n,m]=[12,8]$, $\sigma_{s}=2.14$ $\textrm{\AA}$ and $\varepsilon_{s}=6.36$ kJ/mol for hydrogen atoms.  
The parameter $\eta$ is varied between $1$ and $5.25$ to scan a broad range of hydrophilicities. This would correspond, for real polar sites, to a variation of site density on the solid surface. The interaction potential with $\eta=1$ is shown in Fig.~\ref{HBp}. This interaction potential is inspired by similar potentials calibrated in the case of atomically structured solids to model H-bonding. Yoshida et al. \cite{yoshida_molecular_2014} proposed the potential described by Eq. [\ref{equation29}] to model the interatomic interactions between the atoms of the first solid layer and the water molecules using a parameter $\eta=2.34$.

\section{Results}

Molecular dynamics simulations were performed using LAMMPS  \cite{plimpton_fast_1993}. We considered the strategy above to determine $\tau$ for the prototypical LJ fluid (parameters $\varepsilon$ and $\sigma$).     
Full details regarding the molecular simulations can be found in \emph{Materials and Methods} and \emph{Supplementary Information}.

\begin{figure}[ht]
\centering
\includegraphics[width=8.6cm]{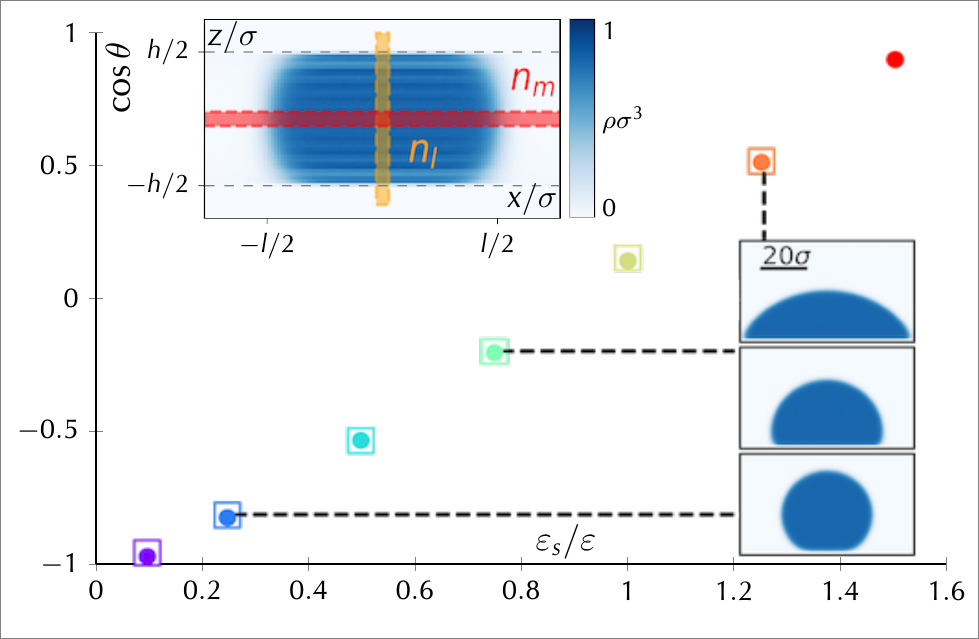}
\caption{Contact angles $\theta$ obtained for a Lennard-Jones fluid from Eq. [\ref{Eq9}] for large confinements (filled circles) or from shape regression for large hemicylinders (empty squares, right insets, see \emph{Supplementary Information}). Contact angles are plotted according to the ratio of the fluid/solid to the fluid/fluid interaction strengths  $\varepsilon_S/\varepsilon$ (values increasing from purple to red). The top inset shows the density of the structured fluid in blue with the yellow and red areas denoting the regions used to define the confinement $h$ and the meniscus separation $l$
(these regions contain $n_{l}$ and $n_{m}$ molecules, respectively, see text).}
\label{Fig2}
\end{figure}

\begin{figure}[ht]
\centering
\subfigure[]{\label{fig3left}\includegraphics[height=0.57\linewidth, keepaspectratio=true]{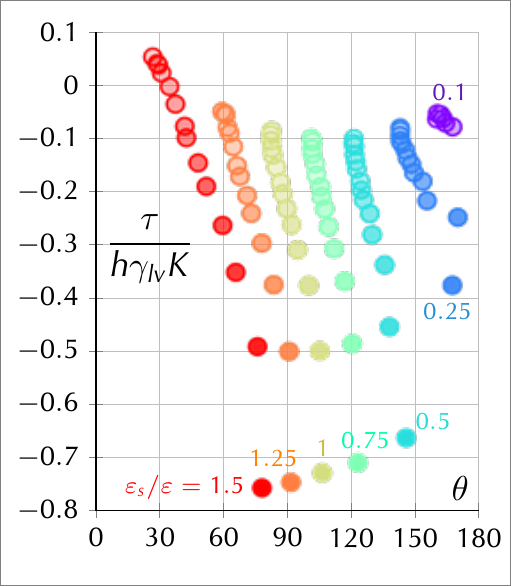}}\hfill
\subfigure[]{\label{fig3right}\includegraphics[height=0.57\linewidth, keepaspectratio=true]{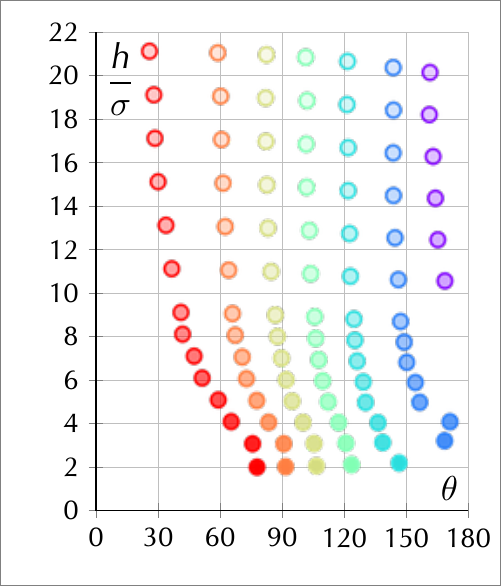}}
\caption{(a) Dimensionless ratio of the line tension $\tau$ over the total force $h\gamma_{sl}K$ due to bulk and surface contributions for several fluid/solid to fluid/fluid interaction strength ratio $\varepsilon_{s}/\varepsilon$ increasing from purple to red. For a given color, the intensity from opaque to transparent indicates increasing  $h$. (b) Dependence of the contact angle $\theta$ on the relative confinement $h/\sigma$ according to the interaction strength ratio (colors).}
\label{Fig3BC}
\end{figure}

As shown in Fig. \ref{Fig2}, $\cos \theta$ determined using Eq. [\ref{Eq9}] follows the expected linear dependence on  $\varepsilon_{s}$ \cite{evans_critical_2016} and, for large $h$, agrees with contact angles measured through the shape regression technique of fluid density maps. {The contact angle as defined by Eq. [\ref{Eq9}] is however not only dependent on the wall/fluid interaction strengths but also on other physical and geometrical parameters such as the confinement (Fig. \ref{fig3right}). This variation emerges from the dependence of solid/vapor and solid/liquid surface tensions on the separation $h$ due to solid-fluid-solid interactions and adsorption effects (see \emph{Supplementary Information}). In particular, solid-fluid-solid interactions induce a significant increase of the contact angle when perfectly wetting surfaces are brought together.}

\begin{figure}[h!]
\centering
\includegraphics[width=8.6cm, keepaspectratio=true]{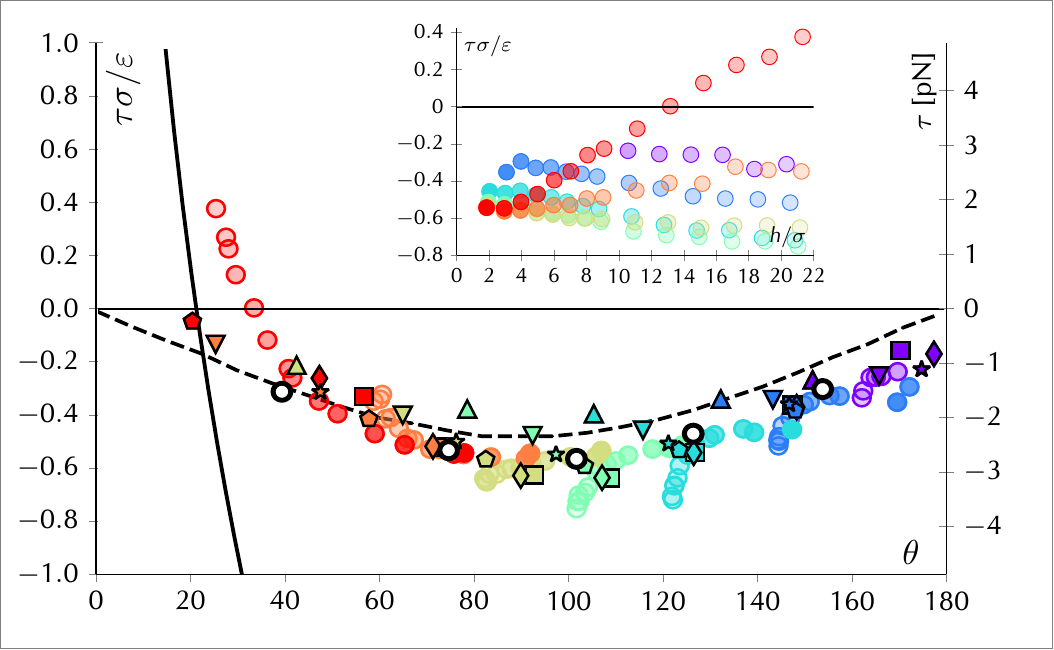}
\caption{Line tension $\tau$ for a LJ fluid confined in slit pores of width $h$ ranging from 2 to 21 $\sigma$. For a given $h$, different interaction strengths $\varepsilon_{s}$ are considered (same color code as in Figs. \ref{Fig2} and \ref{Fig3BC} ). For a given color, the intensity from opaque to transparent indicates increasing  $h$ (see inset showing $\tau$ as a function of $h$ for different $\varepsilon_{s}$).  Different $T$ were considered: $k_{\textrm{B}}T/\varepsilon=0.7~\square, 0.75~\Diamond, 0.85~\pentagon, 0.9~\bigstar, 0.95~\triangledown$, and $1.0~\vartriangle$. The dashed line represents a nonlocal DFT model in a wedge \cite{weijs_origin_2011}. The solid line corresponds to the Interface Displacement Model at the wetting transition \cite{joanny_role_1986}. Simulations with $\sigma_{s}=\sigma$ lead to a similar behavior (black circles). Right axis: same $\tau$ but in real units using LJ parameters for argon ( $\sigma = 3.4$ \AA, $\varepsilon/k_b$ = 120 K). }
\label{Fig3}
\end{figure}

Different wall/fluid interaction strengths $\varepsilon_{s}$, confinements $h$ and temperatures $T$ were considered to probe the value of the line tension $\tau$ on a broad range of parameters for the LJ fluid. 
{At the higher considered temperature, of the order of  0.7 times the critical temperature, the error on the vapor pressure $P_{v}$ given by the perfect gas law is smaller than 5\% \cite{Schlaich2019}.}
We checked for each parameter set, that the magnitude of $\tau$ is significantly larger than the fluctuations on each of the two terms used in the difference in Eq. [\ref{Eq4}] to extract $\tau$ (see Supplementary Information and Fig. S3).  Moreover, as shown in Fig. \ref{fig3left}, in most considered cases $\tau$ is larger than 10\% of the last term of Eq. [\ref{Eq4}] which fully legitimate its measurement from the difference between the two terms of Eq. [\ref{Eq4}].  Among the various considered cases, an important variability in the line tension values is observed. However, as shown in Fig. \ref{Fig3}, all line tension values almost collapse on a single master curve when plotted according to the contact angle $\theta$ extracted for each set of parameters. {Despite the large range of values explored for each parameter the limited departure from this general trend is particularly unexpected. The collapse of $\tau$ on a single curve for different $\varepsilon_{s}$, $\sigma_{s}$, $T$ and $h$ suggests that, to first order, the line tension can be estimated using $\theta$ only. 
This parameterization is possible when using the zero adsorption definition for $h$ and the corresponding definition for $\theta$ given in Eq. [\ref{Eq9}]. }

The line tension plotted according to the contact angle demonstrates a non monotonic behavior with a minimum negative value for $\tau$ around $\theta=90$\textdegree\xspace   and a divergence towards positive values close to the wetting transition. This  behavior makes the bridge between nonlocal Density Functional Theory (DFT) calculations \cite{weijs_origin_2011} (dashed line in Fig. \ref{Fig3}) and Interface Displacement Model (IDM) by Joanny and de Gennes \cite{joanny_role_1986} (solid line in Fig. \ref{Fig3}) at the wetting transition. At the dewetting transition $\theta \to 180$\textdegree\xspace, the vapor wedge separating the liquid from the solid vanishes together with its contact line and its associated free energy $\tau$. For intermediate solid/fluid interaction strengths $0\text{\textdegree\xspace} < \theta < 180\text{\textdegree\xspace}$, the line tension is driven by a combination of microscopic effects that arise from the competition between the different molecular interactions but also potentially from fluid layering at interfaces. Here, analytic sums over molecular interactions do not allow estimating $\tau$ as it corresponds to a free energy contribution that also includes an entropy term \cite{getta_line_1998}. For $\theta =90$\textdegree\xspace, $\tau$ can be seen as a correction of the liquid-vapor surface tension which accounts for the progressive vanishing of pressure anisotropy close to the solid surface. This reduced pressure anisotropy lowers the free energy cost related to the liquid-vapor interface so that $\tau$ is negative (as discussed in \emph{Supplementary Information}, $\tau(90\text{\textdegree\xspace})  \sim -\gamma_{lv}\sigma = -0.74\varepsilon/ \sigma$).
At the wetting transition $\theta \rightarrow 0$\textdegree\xspace, liquid-vapor and solid-liquid interfaces become parallel and close to each other in the vicinity of the contact line. In this configuration, the line tension $\tau$ is mainly controlled by the disjoining pressure which arises from the interaction between these two interfaces.
 In this case, the IDM \cite{joanny_role_1986} predicts that the line tension diverges as $\tau = \gamma_{lv}a (\ln (1/\theta)-1)$ with $a$ a solid-fluid interaction length (see \emph{Supplementary Information}). The molecular simulation data in Fig. \ref{Fig3} are consistent with the divergence predicted at the wetting transition by the IDM (solid line in Fig. \ref{Fig3}).
 
\begin{figure}[ht]
\centering
\subfigure[]{\label{phibs}\includegraphics[width=\linewidth, keepaspectratio=true]{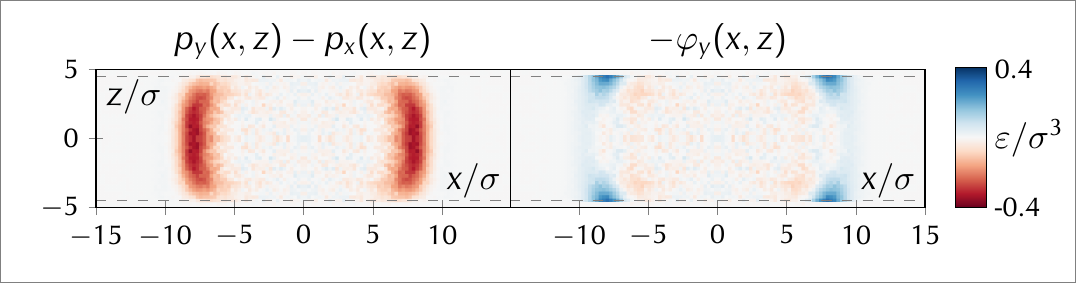}}
\subfigure[]{\label{phiz}\includegraphics[width=\linewidth]{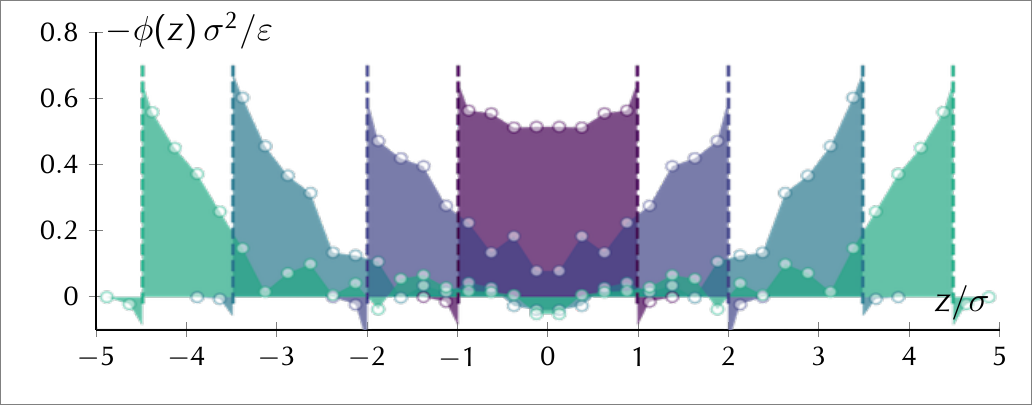}}
\caption{(a) Local pressure anisotropy $p_{x}-p_{y}$ and excess free energy $\varphi(x,z)$ for a LJ fluid confined in a slit pore. The solid-fluid interaction strength $\varepsilon_s$ was chosen so that $\theta = 96$\textdegree\xspace. (b) Distribution of the excess free energy $\phi(z)$ obtained as the integral along $x$ of $\varphi(x,z)$ . Each color plot corresponds to a different $h$ which increases from purple to green. The colored areas, which represent the integral of the excess free energy along the $z$ direction, correspond to $2\tau$. }
\label{Fig4}
\end{figure}

For wetting surfaces $\theta < 90$\textdegree\xspace, the effect of confinement on line tension provides evidence for the limitation of the scaling methodology based on Eq. [\ref{Eq1}] (inset of Fig. \ref{Fig3}). Line tensions are sometimes assumed to correspond to the curvature dependence of the liquid-vapor surface tension described through the Tolman length \cite{kanduc_going_2017,das_contact_nodate}. Yet, a major difference between effects arising from the stress anisotropy close to the contact line and curvature effects lies in the the spatial distribution of the excess free energy. While $\tau$  corresponds to effects localized in the vicinity of the contact line, curvature effects are distributed over the whole liquid-vapor interface (meniscus). 
Fig. \ref{phibs} shows an estimate of the distribution of local stress anisotropy $p_y - p_x$ and local excess free energy  $\varphi(x,z)$ with respect to the bulk and surface value  for a representative example $\theta \sim 90$\textdegree\xspace (stresses were calculated using the Irving-Kirkwood convention \cite{irving_statistical_1950}). Fig. \ref{phiz} shows that the excess free energy $\phi(z)$, obtained from the integration of  $\varphi(x,z)$ along the x direction, is strongly localized close to the triple lines, therefore supporting the line tension concept {(see \emph{Supplementary Information} for computation details)}.

\begin{figure}[ht]
\centering
\includegraphics[width=8.6cm, keepaspectratio=true]{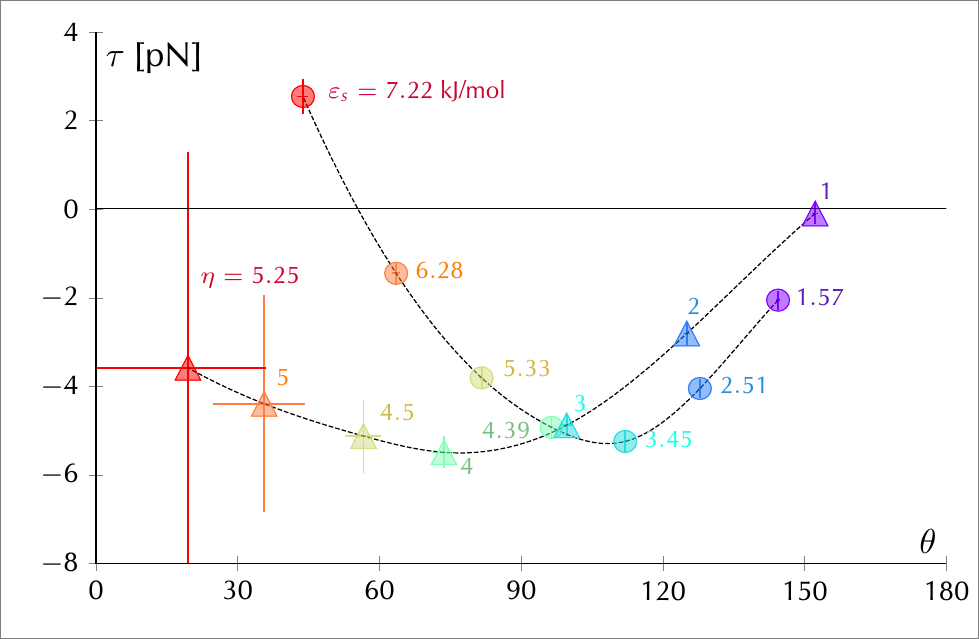}
\caption{Line tension $\tau$ for water confined at room temperature in a slit pore of a width $h =2.6$~nm between dispersive (circles) and hydrogen bonding (triangles) surfaces. For dispersive surfaces, the hydrophilicity is controlled by the solid-liquid interaction strength $\varepsilon_{s}$. For hydrogen bonding surfaces, the hydrophilicity is controlled by the parameter $\eta$ (see \emph{Materials and Methods}). }
\label{Fig5}
\end{figure}

The novel strategy presented here was also used to determine the line tension $\tau$ for water (SPC/E water model) in the vicinity of structureless solid surfaces which interact with water through dispersive interactions or through homogeneously distributed hydrogen bonding \cite{yoshida_molecular_2014} (see \emph{Materials and Methods} and \cite{yoshida_molecular_2014}). Wall-induced polarization effects, ignored with the first type of interaction, are taken into account with the second type of interaction which mimics proton-acceptor sites of real surfaces. 
Fig. \ref{Fig5} shows  $\tau$ for water as a function of $\theta$ for both solid surfaces to investigate the effect of the molecular interactions at play. The two data sets feature a non monotonic behavior similar to that observed for the Lennard-Jones fluid. In both cases $\tau$ is minimum around $90$\textdegree\xspace and vanishes at the dewetting transition $\theta \rightarrow 180$\textdegree\xspace. This general non monotonic behavior suggests that the unifying formalism, described earlier for the LJ fluid, can be used for various fluids and solid/fluid interactions. 

However, an important difference is observed at the wetting transition $\theta \rightarrow 0$\textdegree\xspace. The line tension $\tau$ for water close to dispersive surfaces becomes positive and diverges while it converges to a finite, seemingly negative, value for hydrogen-bonding surfaces. In view of the difference between the two interaction types, it is tempting to qualitatively relate such a behavior to the theoretical predictions of Indekeu \cite{indekeu_line_1992,joanny_role_1986} in the framework of the IDM. In the dispersive case, characterized by $z^{-3}$ attractive interactions between the solid and oxygen atoms of water, like for the LJ fluid, a divergence of the line tension at wetting is predicted. In the hydrogen-bonding case, water interactions with the solid differ in terms of oxygen/solid and hydrogen/solid interactions. Attractive interactions of oxygen and hydrogen atoms with solid plates scale respectively as $z^{-6}$ and $z^{-8}$ (see Fig.~\ref{HBp}). In both cases, these interactions decay faster than those corresponding to Van der Waals interactions (see Fig.~\ref{HBp}) so that they can be referred to short-range interactions according to Indekeu's formalism. The IDM model predicts for such short-range fluid/solid interactions that the line tension converges towards a finite and positive value. The fact that a finite line tension is measured at the wetting transition in the hydrogen bonding case suggests that it is indeed mainly controlled by these so called short-range interactions when $\theta \rightarrow 0$\textdegree\xspace. 

The line tension tension in the case of water confined between either of the two types of surfaces is found to be about $-5$ pN for a contact angle close to $90$\textdegree\xspace. This value, measured at room temperature, is comparable with the value of $-11$ pN extracted by Tinti et al from out of equilibrium simulation of bubble nucleation within a hydrophobic cylindrical nanopore corresponding to a contact angle of $119$\textdegree\xspace \cite{Tinti2017}. The similarity between these two line tension values is remarkable given the strong difference between the two numerical approaches. Even more striking is the proximity of these values with the experimental value of $-25$ pN deduced from the measurement of the extrusion pressure of water out from hydrophobic nanopores having a similar contact angle \cite{guillemot_activated_2012}.

\section{Conclusion}

This study sheds light on the concept of line tension. It unravels a generic behavior, with in the case of dispersive fluids, a main dependence on the contact angle only. Using a novel strategy relying on a mechanical measurement at the molecular scale, our data for the Lennard-Jones fluid and water provide robust line tensions which depend on the wetting properties of the solid surface by the liquid phase. Our line tension values are  found to be consistent with a series of theoretical, numerical and experimental data. The generic behavior emerging from our data,  established on a full range of contact angles, allows unifying the different, sometimes conflicting, pictures in the literature. Far from the wetting transition, the line tensions inferred from our  analysis suggest that this concept is robust down to the molecular scale with a simple dependence on confinement, temperature and solid-fluid interaction encompassed in the contact angle. For a contact angle about $90$\textdegree\xspace, the line tension for water at room temperature is consistent with that inferred from {out of equilibrium numerical and experimental measurements} based  on water extrusion from hydrophobic porous materials. {Beyond measuring line tension values, the computational approach is particularly useful to identify the molecular structures, which are not accessible experimentally, such as layering, adsorption, presence of chemical groups such as OH$^{-}$ responsible for the measured line tension.} 
While the present study is limited to a single liquid/vapor system at equilibrium on unstructured solid surfaces, additional physical features are to be expected for structured and thermalized surfaces, fluid mixtures, and out of equilibrium processes such as bubble nucleation.

\begin{acknowledgements}
This work was supported by the French Research Agency (ANR TAMTAM 15-CE08-0008 and LyStEn 15-CE06-0006). We thank E. Charlaix for fruitful discussions and NVIDIA Corporation for the donation of a Tesla K40 GPU used for this research.
\end{acknowledgements}


\bibliography{bibliography}



\section*{Supplementary material}

# Probing the concept of line tension down to the nanoscale: Supplemental Information

Romain Bey, Benoit Coasne,* and Cyril Picard*

*Univ. Grenoble Alpes, CNRS, LIPhy, 38000 Grenoble, France*

E-mail: benoit.coasne@univ-grenoble-alpes.fr; cyril.picard@univ-grenoble-alpes.fr

## 1 Thermodynamical equilibrium expressed as a function of global geometry and forces

Considering the system simulated in the main article, the global parameters $N$, $L_x$, $L_y$, $h_s$ and $T$ are imposed in each simulation, and are therefore the set of state variables defining the global thermodynamic equilibrium and its associated free potential $F$ (here $h_s$ stands for the imposed distance between the origins of the potentials defining each solid wall). The thermodynamic identity corresponding to this ensemble writes:

$$dF = \Sigma_x dL_x + \Sigma_y dL_y + \Sigma_z dh_s + \mu dN - SdT \tag{1}$$

with $S$ the entropy, $\Sigma_\alpha$ with $\alpha = x,y,z$ the force exerted by the system on the box boundary in the $\alpha$ direction and $\mu$ the uniform chemical potential at equilibrium. $\Sigma_x$ and $\Sigma_y$ (Figs. 1A and 1B) can be measured through the virial theorem. $\Sigma_z$ (Fig. 1C) can be directly measured as the average force exerted on each solid wall in the direction $z$.

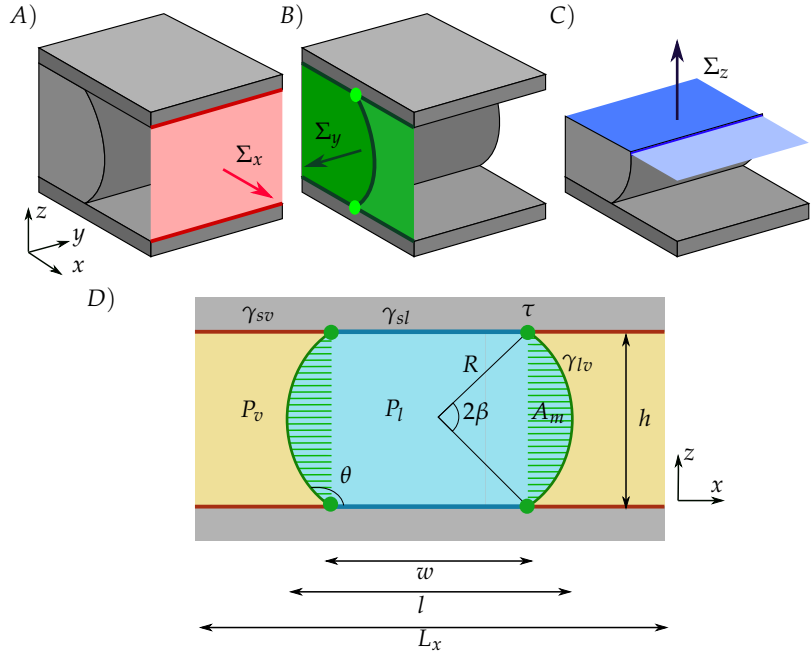

Figure 1: A) Force $\Sigma_x$ in the $x$ direction due to solid-vapor surface tension $\gamma_{sv}$ (red) and vapor pressure (pink). B) Force $\Sigma_y$ due to liquid and vapor pressures $P_l$ and $P_v$ (medium green), solid-liquid, solid-vapor and liquid-vapor surface tensions $\gamma_{sl}$, $\gamma_{sv}$, $\gamma_{lv}$ (dark green) and line tension $\tau$ (green dots). C) Vertical force $\Sigma_z$ due to liquid and vapor pressures (medium blue) and liquid-vapor surface tension (dark blue). D) Force in the $y$ direction $\Sigma_y$ decomposed in $\gamma_{sv}$ and $P_v$ contributions (orange), $\gamma_{sl}$ and $P_l$ contributions (blue) and meniscus contributions $\tau$, $\gamma_{lv}$, and $P_l - P_v$ (green). The angle $\beta$, the radius of curvature $R$ and the area of meniscus in the $xz$ plane $A_m$ are the geometric parameters defining the meniscus.

The virial theorem expresses the free energy derivative $\Sigma_\alpha = \partial F/\partial L_\alpha$ as a function of atomic positions. In the case of a purely fluid system enclosed in a periodic box and interacting through a pair additive interatomic potential, the virial theorem applied in $x$ direction writes:

$$\Sigma_x = -\frac{1}{L_x}\left\langle Nk_B T + \sum_{i,j}^N f_{ij}^x x_{ij} \right\rangle \quad (2)$$

with $f_{ij}^x$ the force applied by atom $j$ on atom $i$ in $x$ direction, and $x_{ij} = x_i - x_j'$ with $x_j'$ the $x$ coordinate of the image through periodic boundary conditions of atom $j$ that is the closest to atom $i$ (minimum image convention). The first term on right hand side of Eq. [2] accounts for the entropic change upon volume modification at constant temperature, and the second term is related to the energetic cost of modifying inter-particle distances.

In the case of a planar interface between a solid and a fluid, different virial expressions are available. If one considers, as it is done in this article, the stretching of both solid and fluid phases, Eq. [2] applied to both solid and fluid particles measures a combination of fluid pressure, solid elastic force and surface stress. If one considers the stretching of the fluid phase on a solid planar surface that features no tangential heterogeneity (the solid is an external field applied on fluid atoms that depends only on their distance to the solid plate), Eq. [2] applied to fluid particles only measures a combination of fluid pressure and solid-fluid surface tension. For water simulations, we follow Thompson et al who give the expression of virial theorem taking into account long range electrostatic interactions and many-body potentials[1].

In addition to Eq. [1] expression, in the case of solid-liquid-vapor systems such as the one considered in this article, the free energy $F$ can equally be divided into contributions related to volumes, surfaces and lines (Gibbs modeling[2]):

$$dF = -P_l dV_l - P_v dV_v + \gamma_{sl} dA_{sl} + \gamma_{sv} dA_{sv} + \gamma_{lv} dA_{lv} + \tau dl_{slv} + \mu dN - SdT \quad (3)$$

with $s$, $v$ and $l$ referring to the solid, vapor and liquid, $P_k$ and $V_k$ the pressures and volumes of bulk phases ($k = l, v$), $\gamma_{kk'}$ and $A_{kk'}$ the surface tensions and areas ($kk' = sv, sl, lv$), $\tau$ and $l_{slv}$ the line tension and length of the triple line.

An important, although counter-intuitive, aspect of line tension $\tau$ as defined through Eq. [3] is its dependence on the definition of geometric parameters (confinement $h$ and contact angle $\theta$ in the geometry considered hereafter, see Fig. 1D). These parameters can be defined in various ways at the molecular scale where surfaces are diffuse by nature. Each definition will therefore lead to different values of line tension $\tau$ and the line tension $\tau$ is called a parametric quantity (as opposed to intrinsic quantities such as $\Sigma_m$ which does not depend on geometric parametrization, see Eq. [6] of the main article). The total free energy $F$ of a system is an intrinsic quantity, which reflects the energy that can be extracted reversibly from it. The subdivision of this total free energy in different contributions necessitates to chose a definition for the position of the interfaces. This is for instance the case for a single interface that separates two bulk phases having different free energy densities $f_v^1$ and $f_v^2$. The surface free energy $\gamma_{12}$ depends on the interface position $z_{12}$ along its normal direction. This dependence is expressed from the notional derivative[3] of the surface energy with respect to the position of the interface $\delta\gamma_{12}/\delta z_{12} = f_v^2 - f_v^1$. This is the case for a solid-fluid interface where the solid free energy is described through its elastic energy and the fluid free energy through its pressure. This dependence of surface tension on surface definition is also found in the case of a curved interface between two fluid phases at different pressures.

## 2 Definition of interfaces position at the microscopic scale

Interfaces between phases are diffuse at the molecular scale, and the location of the separating surfaces depends on conventions. In this article, we adopt the zero adsorption convention, that consists in choosing the surface locations that suppress the excess of fluid at the solid-liquid and liquid-vapor interfaces. We therefore compute the integrated densities $n_m = \int dx \rho(x, z=0)$ and $n_l = \int dz \rho(x=0, z)$ in horizontal and vertical slices far from the triple line (red and yellow regions in Fig. 3 of the main article). The location of the solid surfaces is then defined through $h = n_l/\rho_l$ with $\rho_l$ the density of a homogeneous liquid phase at the same temperature, and the separation of the meniscii $l$ is defined as $l = [n_m - \rho_v L_x]/[\rho_l - \rho_v]$ with $\rho_v$ the vapor density in the middle of the pore far from the meniscii.

The definitions of $h$ and $l$ require the knowledge of the liquid bulk density $\rho_l$ as a function of pressure $P$ and temperature $T$. $\rho_l$ is measured from NPT simulations containing 10648 atoms of a homogeneous Lennard-Jones liquid phase (Fig. 2A). For the pressure and temperature ranges considered here, as shown in Fig. 2C, the relative variations in $\rho_l$ with $P$ are weak and slightly impact the definition of $h$. For the definition of $h$, we use the bulk density $\rho_l$ at the saturation pressures. The definition on the contact angle $\theta$, requires moreover the knowledge of the vapor pressure $P_v$. This pressure was measured for each system from the vapor density in the middle of the dry pore region using the ideal gas law, $P_v = \rho_v k_B T$. Due to the finite size of the volume control used for such calculations, large fluctuations are observed in the number of molecules as a function of time but we estimated that the error bar over $P_v$ is less than 3%.

The liquid-vapor surface tension $\gamma_{lv}$ has to be known to extract the line tension $\tau$ and to define the contact angle $\theta$ (Eqs. [6] and [9] of the main article). For the Lennard-Jones fluid, we simulate a liquid slab containing 3375 atoms in a periodic box of dimensions $[15\sigma, 15\sigma, 80\sigma]$ (Fig. 2B). The liquid-vapor surface tension $\gamma_{lv}$ is determined from the difference between the normal and tangential pressures that are measured from the stresses in the vertical and horizontal directions (inset Fig. 2C). We simulate this system for a duration of $10^7 \Delta t$ at the temperatures $k_B T/\varepsilon = 0.7, 0.8, 0.9, 1.0$. For the case $Tk_B/\varepsilon = 0.8$, we run a longer

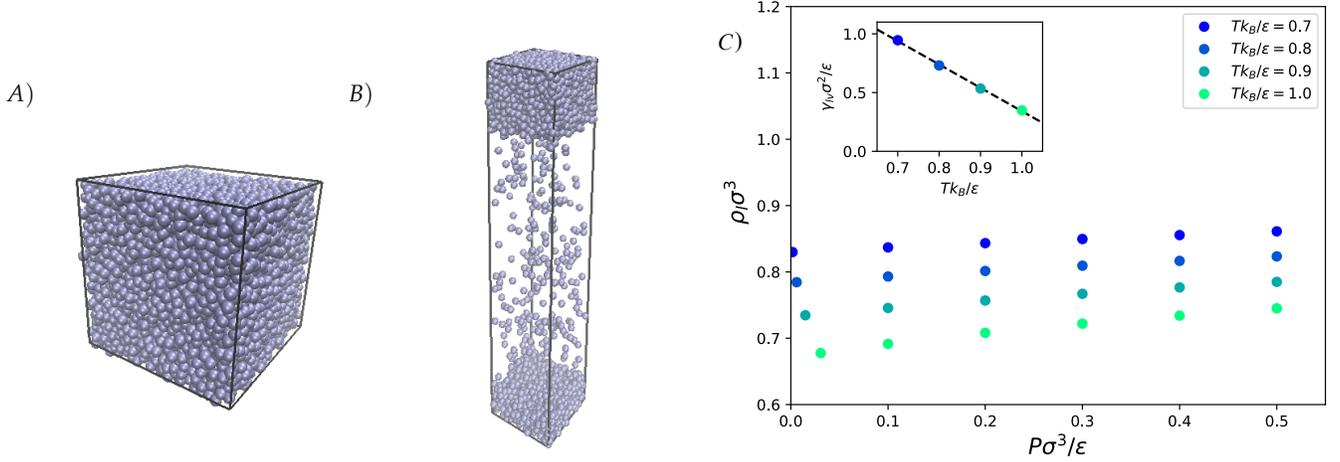

Figure 2: A) Homogeneous Lennard-Jones liquid in a periodic simulation box. B) Lennard-Jones liquid-vapor interfaces in a periodic box. C) Liquid density $\rho_l$ at various pressures $P$ and temperatures $T$ of a Lennard-Jones fluid. Inset: liquid-vapor surface tension $\gamma_{lv}$ as a function of temperature approximated by a linear fit (dashed line). All quantities are given in reduced units with respect to the LJ parameters $\sigma$ and $\varepsilon$.

simulation of total duration $10^8 \Delta t$. For the temperature range considered here the liquid-vapor surface tension $\gamma_{lv}$ decreases linearly with temperature as shown in the inset of Fig. 2C. In the case of water simulations, we simulate a liquid film containing 3456 molecules in a periodic simulation box of dimensions [4 nm, 4 nm, 30 nm] for a duration of 50 ns at the temperature of 300 K.

## 3 Derivation of internal equilibrium relations

The liquid volume writes $V_l = L_y h w + 2 L_y A_m(h, \theta)$, where the shaded area $A_m$ in Fig. 1D is a function of confinement $h$ and contact angle $\theta$ only. The surface $A_m$ is an algebraic quantity that is positive for $\theta > \pi/2$ ($P_l > P_v$) and negative for $\theta < \pi/2$ ($P_l < P_v$).

As the molecular dynamics simulation is run in the NVT ensemble, the total volume $V = V_l + V_v$ is constant. The area $A_{sl} = 2 w L_y$ and $A_{sv} = 2(L_x - w) L_y$ are independent of $\theta$ while the area $A_{lv} = 2 \mathcal{L} L_y$ with $\mathcal{L} = 2 \beta R = \beta h / \sin \beta$ the arc length of the liquid vapor interface in the $xz$ plan, is independent of $w$. At equilibrium, each partial derivative of $F$ with respect to $\theta$ and $w$ is null. The derivative with respect to $w$ gives

$$\left. \frac{\partial F}{\partial w} \right|_{h, \theta} = L_y h (P_v - P_l) + 2 (\gamma_{sl} - \gamma_{sv}) L_y = 0$$

which leads to:

$$2(\gamma_{sl} - \gamma_{sv}) - (P_l - P_v) h = 0 \qquad (4)$$

The derivative with respect to $\theta$ writes

$$\left. \frac{\partial F}{\partial \theta} \right|_{h, w} = 2 L_y (P_v - P_l) \frac{\partial A_m}{\partial \theta} + 2 \gamma_{lv} \frac{\partial \mathcal{L}}{\partial \theta} L_y = 0 \qquad (5)$$

Using the expression of the algebraic area $A_m$

$$A_m = (\beta - \cos \beta \sin \beta) h^2 / (4 \sin^2 \beta) = \frac{h}{4} \left( \frac{\mathcal{L}}{\sin \beta} - \frac{h}{\tan \beta} \right)$$

and noting that $\beta = \theta - \pi/2$ one gets:

$$\frac{\partial \mathcal{L}}{\partial \theta} = \frac{\partial \mathcal{L}}{\partial \beta} = \frac{h}{\sin \beta} - \frac{\mathcal{L}}{\tan \beta}$$

$$\frac{\partial A_m}{\partial \theta} = \frac{\partial A_m}{\partial \beta} = \frac{h}{4} \left( \frac{\partial \mathcal{L}}{\partial \beta} \frac{1}{\sin \beta} - \frac{\mathcal{L}}{\sin^2 \beta} \cos \beta + \frac{h}{\sin^2 \beta} \right) = \frac{h}{2 \sin \beta} \frac{\partial \mathcal{L}}{\partial \theta}$$

3

As a result, replacing the partial derivative of $A_m$ in Eq. [5]

$$2\gamma_{lv} \sin\beta + (P_v - P_l)h = 0$$

leads to the Young equation when using Eq. [4] to express $P_v - P_l$:

$$\gamma_{sv} - \gamma_{sl} - \gamma_{lv}\cos\theta = 0 \tag{6}$$

## 4 Uncertainty on line tension

The line tension is obtained from the difference of two terms as described by Eq. [6] of the main article. The uncertainty is

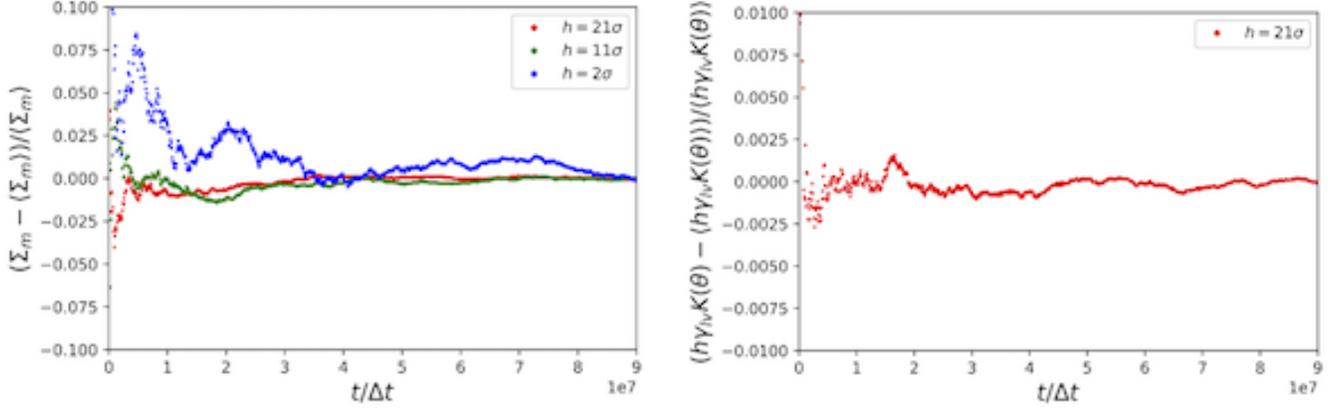

Figure 3: Left: fluctuations of the measured average meniscus free energy $\Sigma_m$ as a function of simulation duration $t$, for $\varepsilon_s/\varepsilon = 1$ ($\theta = 90°$) at three different confinements $h$ (colors). Right: Fluctuations of the measured average $h\gamma_{lv}K(\theta)$ as a function of simulation duration $t$, for $\varepsilon_s/\varepsilon = 1$ ($\theta = 90°$) and $h/\sigma = 21$ ($\gamma_{lv}$ is measured independently in an additional simulation of a flat liquid-vapor interface and its fluctuations are not considered here).

estimated from the relative fluctuations over each term.

As an illustration, Fig. 3 shows these relative fluctuations according to the number of time steps for each term for a contact angle of 90° and a temperature corresponding to $k_B T/\varepsilon = 0.8$. From the typical error bars, the sensitivity on line tension is of the order of $0.05\varepsilon/\sigma$, that is with argon parameters of the order of $2 \times 10^{-13}$ N.

## 5 Impact of confinement

### 5.1 Dependence of solid/fluid interfaces position on confinement

As explained above, we have chosen a definition of the confinement $h = n_l/\rho_l$ that corresponds to the zero adsorption condition. Another common definition consists in choosing the origin of the solid potential $z_s$ as the definition of the surface, leading to the confinement $h_s = z_2 - z_1$. We show in Fig. 4 the difference between these two definitions. This difference significantly depends on the confinement $h$ in the case of very hydrophobic substrates due to high surface compressibility (see Fig. 5C for the Lennard-Jones fluid).

### 5.2 Dependence of contact angle on confinement

The parameter defined by Eq. [9] of the main article writes:

$$\xi = \frac{\gamma_{sv} - \gamma_{sl}}{\gamma_{lv}} \tag{7}$$

where the solid/vapor and solid/liquid surface tensions are effective surface tensions that vary with $h$ due to solid-fluid-solid interactions and with the chemical potential $\mu$ due to adsorption effects. In the range $-1 < \xi < 1$ one can define a contact angle $\theta$ such that $\cos\theta = \xi$. Perfect wetting associated to positive spreading coefficient correspond to situation where $\xi > 1$ while fully non wetting situation correspond to situation where $\xi < -1$. The pressure dependence of the liquid-vapor surface tension $\gamma_{lv}$ is neglected since it has been proven to be small when using the zero adsorption definition to locate the liquid-vapor interface[4].

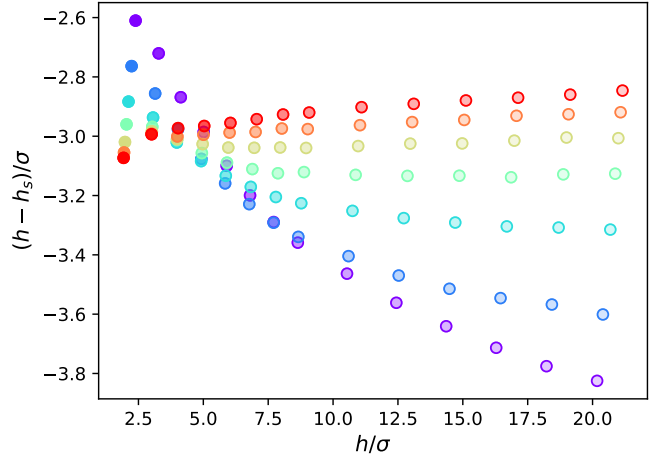

Figure 4: Difference between the confinement defined with respect to the zero adsorption surfaces $h$ and the confinement defined with respect to the origins of the solid potentials $h_s = z_2 - z_1$, as a function of the confinement $h$ for Lennard-Jones simulations. The colors correspond to different interaction parameters $\varepsilon_s$ which increase from purple to red (very hydrophobic to very hydrophilic).

Fig. 5 A and B show the contact angle $\theta$ computed at various hydrophilicities $\varepsilon_s$, confinements $h$ and temperatures $T$ for the LJ simulations. The strong variation of $\cos\theta$ with $h$ for a given $\varepsilon_s$ underlines the limitations of the classical size scaling technique that assumes $\theta$ to be constant in absence of triple line curvature. Confinement-induced effects in the hydrophobic cases are due to an important variation of the fluid structure close to the solid as the pressure is increased. At high pressure, the fluid is pushed closer to the origin of the solid potential, which can be understood as a surface compressibility effect (Fig. 5C). The large compressibility of the solid-liquid surface at the dewetting transition ($\theta \to 0°$) has been extensively considered in another publication[5]. Moreover, the structure of the fluid phase is pressure-dependent, and fluid layering occurs at high pressure so that the free energy associated to the interface is modified. We now consider the hydrophilic cases and focus on the impact of solid-fluid-solid interactions. Assuming a local description of surface free energies, we write the dry-wet equilibrium as $2\gamma_{sv} - P_v h = 2\gamma_{sl} - P_l h$ (Eq. [4]). We call $\gamma_{sk}^\infty$ with $k = v, l$ the surface tension measured for a solid wall interacting with an infinite half space of fluid at the same chemical potential, and $W_{sks}$ the correcting term appearing in the dry-wet equilibrium equation due to nonlocal effects. The dry-wet equilibrium can be rewritten as:

$$2\gamma_{sv}^\infty - P_v h + W_{svs}(h) = 2\gamma_{sl}^\infty - P_l h + W_{sls}(h) \tag{8}$$

The correction terms can we approximated by a purely energetic model, where $W_{sks}$ corresponds to the missing interactions of the solid with the fluid phase:

$$W_{sks} = -2 \int_{h/2}^\infty \rho_k(z) u^1(z) dz \tag{9}$$

with $k = v, l$ and $u^1$ the interacting potential with the bottom solid. The factor 2 in the above equation accounts for the fact that $W_{sks}$ corrects the two surface tensions $\gamma_{sk}$. Focusing on the long range attractive component of the solid-liquid interaction and on the dense liquid phase, we consider $W_{svs} = 0$ and approximate $W_{sls}$ as:

$$W_{sls} = 2\rho_l \int_{h/2}^\infty \varepsilon_s \left(\frac{\sigma_s}{z - z_{s1}}\right)^3 dz \tag{10}$$

$$= \rho_l \frac{\varepsilon_s \sigma_s^3}{(h/2 - z_1)^2} \tag{11}$$

Without the nonlocal interaction term $W_{sls}$, the force $\Sigma_z$ in the $z$ direction writes (see Fig. 1CD):

$$\Sigma_z = L_y \left(2\gamma_{lv} - P_l l - P_v(L_x - l)\right) \tag{12}$$

Inserting the derivative of the correcting energy $W_{sls}$ in the normal force expression Eq. [12], we obtain:

$$\Sigma_z = L_y \left(2\gamma_{lv} - (P_l + \frac{\partial W_{sls}}{\partial z_1}) l - P_v(L_x - l)\right) \tag{13}$$

5

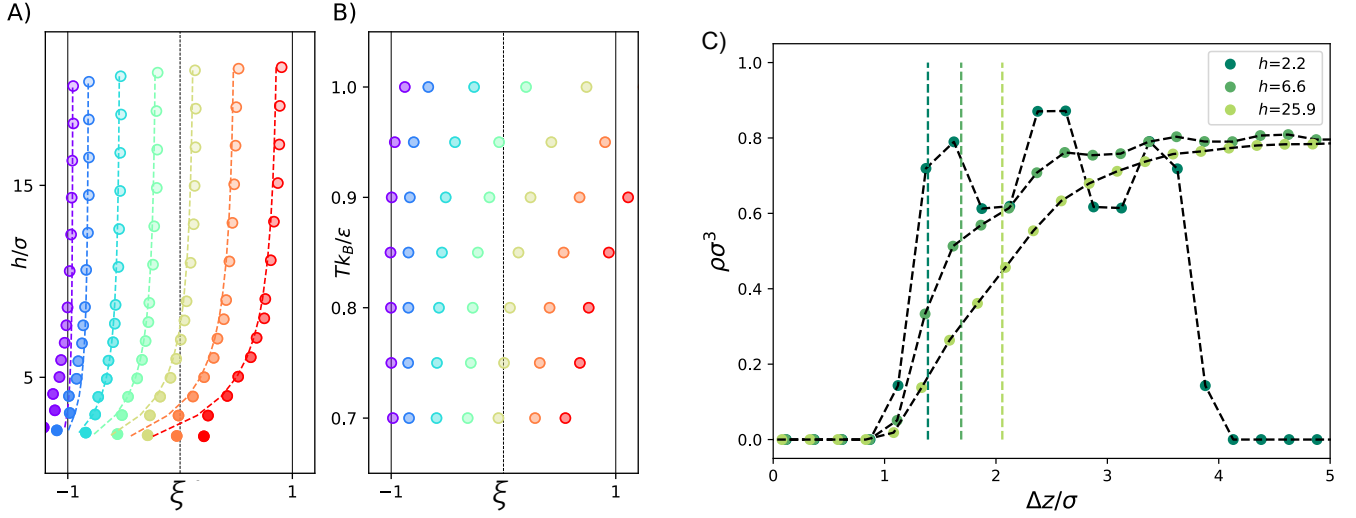

Figure 5: (AB) $\xi$ defined using Eq. [9] of the main article for a LJ fluid. For values beyond the range [-1,1] this parameter should be seen as the spreading ratio $(\gamma_{sv} - \gamma_{sl})/\gamma_{lv}$ at various confinements $h$ (A), temperatures $T$ (B) and solid-fluid interaction parameters $\varepsilon_s$ (colors as defined in the main article). A simple model accounts for confinement effects (dashed line, see text). The simulations at various temperatures correspond to $h/\sigma \approx 9$. (C) Fluid density profile in the liquid phase normal to the wall at different confinements $h$ (green points). These data are for the LJ fluid in contact with the most hydrophobic solid. $\Delta z$ is the coordinate difference $z - z_1$ with $z_1$ the origin of the solid potential. For each confinement, the dashed vertical line indicates the surface corresponding to the zero adsorption condition.

Replacing $\Sigma_z$ by Eq. [13] in Eq. [9] of the main article, the parameter $\xi$ writes:

$$\xi = -\frac{(P_l - P_v)h}{2\gamma_{lv}} - \frac{h}{2\gamma_{lv}}\frac{\partial W_{sls}}{\partial z_1} \tag{14}$$

Inserting Eq. [8] of the Supplemental Information in the last equation and neglecting $W_{svs}$ brings:

$$\xi = \frac{\gamma_{sv}^\infty - \gamma_{sl}^\infty}{\gamma_{lv}} - \frac{W_{sls}}{2\gamma_{lv}} - \frac{h}{2\gamma_{lv}}\frac{\partial W_{sls}}{\partial z_1} \tag{15}$$

Eqs. [11] and [15] are plotted in Fig. 5A taking for $(\gamma_{sv}^\infty - \gamma_{sl}^\infty)/\gamma_{lv}$ the value of $\xi$ measured through Eq. [9] of the main article in the large pore limit. This model accurately describes the variations of $\xi$ and associated contact angle $\theta$ with confinement.

For large $h$, contact angles measured through Eq. [9] of the main article are compared with macroscopic contact angles $\theta^{Macro}$ measured from shape recognition on large hemicylindrical drops on a solid surface (Fig. 3 of the main article). We simulate therefore 24300 LJ particles in a box of dimensions $[200\sigma, 21\sigma, 150\sigma]$ for a total time of $10^7 \Delta t$. The system relaxes for $10^6 \Delta t$ before recording an average density map. We measure the location $z_0$ of the zero adsorption surface in the middle of the drop. We fit a circle on the points of the density map such that $\rho_{fit} = (\rho_l + \rho_v)/2$ and $z - z_0 > 8\sigma$. The contact angle is defined as the angle of the fitted circle with the plane of equation $z = z_0$ (Fig. 6A).

For small $h$, a first remark concerns the parameter $\sigma_s$ in the Lennard-Jones simulations. Apart from five simulations that have been run with $\sigma_s = \sigma$, all the other simulations have used the value $\sigma_s = 2\sigma$. We have chosen this large solid-fluid characteristic length to emphasize the effects due to the long range of dispersive solid-fluid interactions. Eq. [11] indicates that, for a given $\varepsilon_s$, dividing $\sigma_s$ by 2 would lead to a reduction of the confinement effects by a factor 8: the confinement effects are strongly dependent on the ratio $\sigma_s/\sigma$. For water simulations with a dispersive solid, we have chosen $\sigma_s = 3$ Å (approximately equal to the diameter of a water molecule). Fig. 6B shows indeed that the confinement effects are much weaker for water simulations.

### 5.3 Dependence of line tension on confinement with water

Fig. 6C shows the line tension $\tau$ of water confined between some dispersive ($\varepsilon_s = 0.6, 1.05, 1.5$ kcal mol$^{-1}$) and hydrogen-bonding ($\eta = 5$) surfaces at various confinements $h$. A drift with the confinement can be observed until $h \approx 2.5$ nm. For extreme confinements, the hydrogen-bonding case features a discontinuous jump reflecting the contact angle shift visible in Fig. 6B.

6

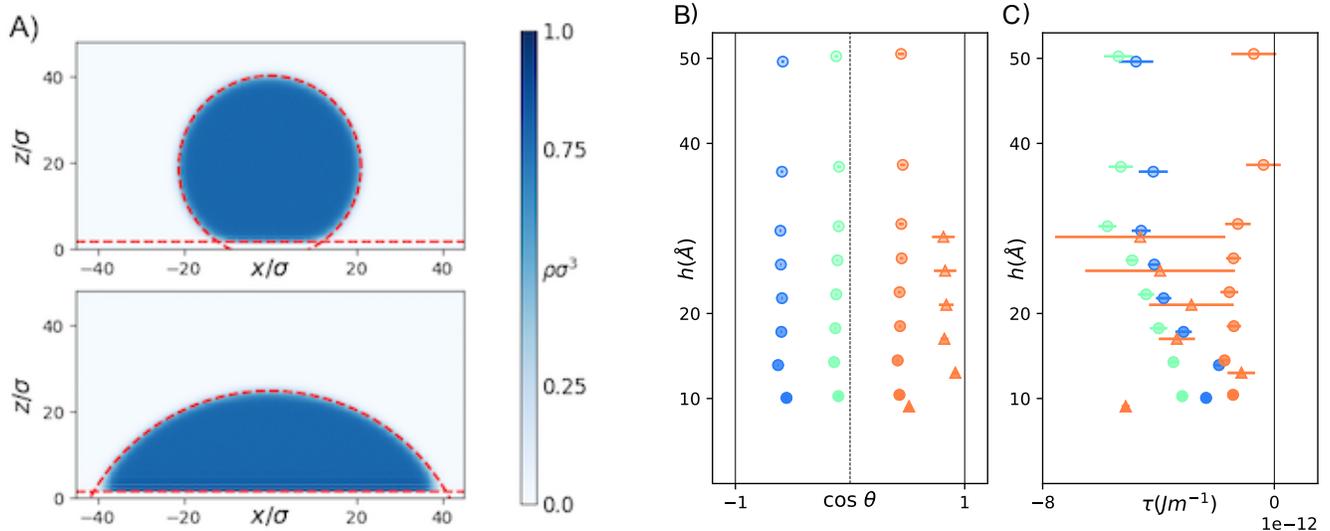

Figure 6: (A) Density maps of large 2D hemicylindrical drops on a solid substrate. The contact angles $\theta$ are defined with respect to fitted circles on the points of intermediate density and with respect to the solide-liquid zero adsorption surface (dashed red lines). (B) Cosine of the contact angle $\theta$ and (C) line tension $\tau$ of water at $T = 300$K, in contact with dispersive (circles) and hydrogen-bonding (triangles) surfaces at various confinements $h$. For the dispersive surface, $\varepsilon_s = 0.6, 1.05, 1.5$ kcal/mol (from blue to orange). For the hydrogen-bonding surface, $\eta = 5$.

## 6 Theoretical prediction for line tension values

### 6.1 Interface Displacement Model

Close to the the wetting transition ($\theta \to 0°$) the long range of the solid-fluid dispersive interactions strongly modifies the liquid wedge geometry so that the liquid-vapor interface in the vicinity of the triple line is bent. Focusing on the long range of the solid-liquid interaction and assuming that liquid-vapor surface tension is purely local, a simple modeling called Interface Displacement Model is available. Joanny and de Gennes showed that the variation of the line tension $\tau$ in this mixed local-nonlocal model (Interface Displacement Model) is given by[6]:

$$\tau = \gamma_{lv} a \left( \ln \frac{1}{\theta} - 1 \right) \tag{16}$$

with $a$ a characteristic length of the solid-liquid-vapor disjoining pressure ($a = \sqrt{A/6\pi\gamma_{lv}}$ with $A$ the Hamaker constant). Extending the estimate of nonlocal energies $W_{sls}$ given by Eq. [11] to a liquid film of width $e$ in contact with solid and vapor phases, we have $W_{slv}(e) \approx \rho_l \varepsilon_s \sigma_s^3 / (2e^2)$ (the factor 2 comes from the fact that only one solid is interacting with the fluid in the solid-liquid-vapor case). Deriving this expression with respect to $e$ leads to the value of Hamaker constant $A = 6\pi\rho_l\varepsilon_s\sigma_s^3$ and the characteristic length:

$$a = \sqrt{\frac{\rho_l \varepsilon_s \sigma_s^3}{\gamma_{lv}}} \tag{17}$$

which has a value of $a \approx 3.6\sigma$ in the case $\varepsilon_s = 1.5\varepsilon$ and $\sigma_s = 2\sigma$. The continuous curve plotted in Fig. 5 of the main article corresponds to Eq. [16] with this value of $a$. This modeling accounts for the liquid-vapor interface bending in the presence of a strong solid-fluid interaction that is long ranged, and is therefore relevant for dispersive surfaces at the wetting transition.

### 6.2 Nonlocal DFT

Another simple modeling consists in neglecting the liquid-vapor surface bending, and on assuming a perfect wedge geometry. If the layering of the fluid induced by the solid is moreover neglected, a numerical computation of a simple DFT model provides the variations of the line tension $\tau$ with the contact angle $\theta$ shown as the dotted line in Fig. 5 of the main article[7]. This modeling is relevant as long as the nonlocal disjoining pressure and its associated liquid-vapor surface bending are negligible.

## 6.3 Estimation of line tension at $\theta = 90°$

We consider the special case of a wetting contact angle $\theta = 90°$. In that case, the geometry is considerably simplified due to the absence of curvature at the liquid-vapor interface. The free energy of the meniscus writes $\Sigma_m = h\gamma_{lv} + 2\tau$. The line tension is the correction accounting for the fact that the meniscus free energy is not exactly equal to $h\gamma_{lv}$, with $\gamma_{lv}$ the liquid-vapor surface tension.

The liquid-vapor surface tension arises as a consequence of the anisotropy of the pressure in the vicinity of the liquid-vapor interface. Close to the solid, fewer fluid particles are involved in the fluid pressure tensor, which leads to a decrease of its anisotropy (Fig. 6A of the main article). This decrease is at the origin of a lower liquid-vapor surface tension in the vicinity of the solid. This effect occurs on a length characteristic of the fluid-fluid interaction, here $\sigma$. The induced correction on the total free energy $h\gamma_{lv}$ can be approximated by $\tau \approx -\sigma\gamma_{lv}$ for each triple line. This estimate is in good agreement with the measured value of $\tau$.

Fig. 6 of the main article considers a dispersive solid interacting with a LJ fluid through a 9-3 potential of parameters $\varepsilon_s/\varepsilon = 0.82$ and $\sigma_s/\sigma = 2$ at four different solid spacings $h_s/\sigma = 5, 7, 10, 12$. The insets represent the pressure anisotropy $p_y - p_x$ (left) and the excess free energy attributed to the line tension $\varphi(x,z)$ (right) in the $h_s/\sigma = 12$ case corresponding to a contact angle $\theta = 96°$. $\varphi(x,z)$ is measured removing the contribution of a flat liquid-vapor interface to the pressure anisotropy: $\varphi(x,z) = p_x(x,z) - p_y(x,z) - \frac{\gamma_{lv}}{\sqrt{2\pi w_0^2}} \exp \frac{-(|x|-x_0)^2}{2w_0^2}$ with $w_0$ and $x_0$ parameters that are fitted in the pore center ($z = 0$) and correspond to the width and position of the interfacial region. Pressure fields are measured using Irving-Kirkwood convention[8].



\end{document}